\documentclass[epj]{svjour}
\usepackage{epsfig}

\newcommand{\equ}[1]{(\protect\ref{#1})}

\begin{document}

\title{Anomalous scaling in the Zhang model}

\author{Romualdo Pastor-Satorras\inst{1} \and Alessandro
  Vespignani\inst{2}}

\institute{Dept. de F{\'\i}sica Fonamental, Facultat de
  F{\'\i}sica, Universitat de Barcelona,
  Av. Diagonal 647, 08028 Barcelona, Spain \and 
  The Abdus Salam International Centre for Theoretical Physics
  (ICTP),  P.O. Box 586, 34100 Trieste,
  Italy}

\date{\today}

\abstract{We apply the moment analysis technique to analyze large
  scale simulations of the Zhang sandpile model. We find that this
  model shows different scaling behavior depending on the update
  mechanism used. With the standard parallel updating, the Zhang model
  violates the finite-size scaling hypothesis, and it also appears to
  be incompatible with the more general multifractal scaling form.
  This makes impossible its affiliation to any one of the known
  universality classes of sandpile models. With sequential updating,
  it shows scaling for the size and area distribution.  
  The introduction of stochasticity into the
  toppling rules of the parallel Zhang model leads to a scaling
  behavior compatible with the Manna universality class.
}

\maketitle

The identification of universality classes is one of the most
important and still open problems in the field of self-organized
criticality (SOC) \cite{jensen98}.  In spite of the great relevance of
this issue, however, it has not been possible until very recently to
clearly discern differences in the critical behavior of the various
SOC models proposed so far.  The situation seems now to have been
settled in the case of the Bak-Tang-Wiesenfeld (BTW) \cite{btw1} and
the Abelian Manna \cite{manna,dhar} models that represent,
respectively, the prototypical examples of deterministic and
stochastic sandpile automata.  In this particular case, recent large
scale simulations \cite{milshtein98,granada,lubeck00,vdmz99} clearly
indicate that Manna and BTW models belong to different universality
classes.

The universality class asset remains still uncertain for the Zhang
model \cite{zhang89}, which is the archetype of all sandpile automata
with continuous variables.  This model has deterministic dynamics like
the BTW model, and in contrast with most other cases, it is {\em
  non-Abelian}.  This means that the stable configurations obtained at
the end of an avalanche depend on the order in which the active sites
are updated. In spite of this essential peculiarity, earlier
simulations of the model \cite{zhang89} placed it in the same
universality class as the BTW model, of which it was supposed to be
the continuous counterpart.  This conclusion was confirmed by the
large scale simulations performed by L{\"u}beck in $d=2$
\cite{lubeck97b}.  On the other hand, Milshtein {\em et al.}
\cite{milshtein98}, analyzing different magnitudes than L{\"u}beck,
arrived at the opposite result in $d=2$, namely, they observed
noticeable differences in the critical exponents.  Finally, the
simulations of Giacometti and D{\'\i}az-Guilera \cite{giacometti98}
provided evidence that, even though the exponents of both models are
similar in $d=2$, they do not coincide in $d=3$. 

Recently, it has been shown \cite{men,tebaldi99}
that the deterministic nature of the BTW model renders its scaling
incompatible with the standard finite-size scaling (FSS) hypothesis,
and induces moreover peculiar non-ergodic effects \cite{vdmz99}. Thus,
it would not be surprising to find similar anomalies in the Zhang
model because of its deterministic nature.  
In this paper we characterize the Zhang universality class by
applying the recently proposed moment analysis technique
\cite{men,tebaldi99}.  In the following we will show that the scaling
of the Zhang model depends very strongly on the updating mechanism
implemented in the simulations, either parallel or sequential. The
Zhang model with parallel updating, as it has been customarily 
defined in the literature, displays a complex scaling behavior  that is not
compatible neither with the standard finite-size scaling (FSS)
hypothesis nor with the multifractal picture \cite{kad89} proposed for
the avalanche distribution of the BTW model \cite{tebaldi99}. On the
other hand, the Zhang model with sequential updating shows 
well defined size and area exponents.  
The origin of the complex behavior of the parallel
Zhang model can  be ascribed to the {\em deterministic} microscopic
dynamical rules of the model. In order to prove this conjecture, we
study a variation of the Zhang model, the stochastic parallel Zhang
model, which exhibits a standard FSS behavior, compatible with the
universality class of the Manna model.

We consider the definition of the original Zhang model \cite{zhang89}.
On each site of a $d$ dimensional hypercubic lattice of size $L$ we
assign a {\em continuous} variable $E_i$, called ``energy''.  At each
time step, an amount of energy $\delta$ is added to a randomly
selected site $j$ ($E_j \to E_j + \delta$).  The quantity $\delta$ is
a random variable uniformly distributed in $[0, \delta_{\rm max}]$
\cite{note1}.  In our simulations we consider the fixed value
$\delta_{\rm max}=0.25$. When a site acquires an energy larger than or
equal to $1$, $(E_k \geq 1$), it becomes active and topples. An active
site $k$ relaxes losing {\em all} its energy, which is equally
distributed among its nearest neighbors: $E_k \to 0$, and $E_{k'} \to
E_{k'}+ E_k/2d$.  Here the index $k'$ runs over the set of all
nearest neighbors of the site $k$. The transported energy can activate the
nearest neighbor sites and thus create an avalanche.  Energy can be
lost only at the boundary of the system (open boundary conditions).
The avalanche stops when all sites in the lattice are subcritical
($E_i<1$).  Given these dynamical rules, it is easy to see that the
Abelian nature of the model depends on the type of updating
implemented. With {\em parallel} updating---parallel Zhang (P-Z)
model--- at each time step $t$ in the evolution, all active sites are
toppled simultaneously, and time is incremented $t \to t + 1$. Since
all the energy of an active site is transferred to its nearest
neighbors, we notice that in a bipartite lattice (such as the
hypercubic lattice here considered) all active sites at a give time
step $t$ reside onto the same sublattice, and that activity
alternates between the dual sublattices in consecutive time steps.
Again, since all the energy is transferred in a toppling, the state of
the active sites in a sublattice at time $t$ is independent of the
order in which the active sites in the dual sublattice were updated at
time $t-1$. On the other hand, with sequential
updating---sequential Zhang (S-Z) model--- at each time step $t$ an
active site is randomly chosen among all the $N_a(t)$ active sites
present at that time. The chosen site is toppled and time is
incremented $t \to t + 1/N_a(t)$. In this case, activity is not
restricted to alternate sublattices, but spreads all over the system.
Depending on the order in which the intermediate list of over-critical
sites is updated, any active site with at least one active nearest
neighbor can end up with an energy $E_i=0$ (if its nearest neighbors
are updated before it) or with energy $E_i>0$ (if it is updated before
its nearest neighbors). In this case the model fully exploits its 
non-abelian characters.

We have analyzed both parallel and sequential Zhang models by
determining their critical exponents \cite{jensen98}.  In the limit of
an infinite slow driving, i.e. the energy addition is interrupted
during the avalanche evolution, defining a complete time scale
separation \cite{vz}, the system reaches a critical stationary state
with avalanches of activity distributed according to power laws.  If
we define the probability distributions $P(x)$ of occurrence of an
avalanche of a given size $s$, time $t$, and area $a$, the FSS
hypothesis \cite{cardy88}, usually assumed in SOC systems, states that
\begin{equation}
  P(x,L) = x^{-\tau_x}{\cal F} \left(\frac{x}{L^{\beta_x}}\right),
  \label{eq:fss}
\end{equation}
with $x=s, t$, or $a$, respectively. If the FSS ansatz is valid, then
the critical exponents $\beta_x$ and $\tau_x$ completely determine the
universality class of the model under scrutiny \cite{notezhang}.

Previous numerical works on the Zhang model
\cite{zhang89,lubeck97b,giacometti98} have most often proceeded
measuring the exponents as the slope in a log-log plot of the density
$P(x,L)$ as a function of the magnitude $x$.  Even though with this
procedure one can determine the exponents within a $10\%$ accuracy,
its performance is affected by the existence of the upper and lower
cutoffs, which render difficult its application.  In this respect, it
is better to use analysis techniques that contain explicitly the
system-size dependency.

The moment analysis technique was introduced by De Menech {\em et al.}
in the context of the two dimensional BTW model~\cite{men}, and its
validity has been extensively checked for both Abelian and stochastic
models~\cite{granada,lubeck00}.  In this me\-th\-od, the $q$-th
moment of a probability distribution on a lattice of size $L$ is
defined by $\left<x^q\right>_L = \int x^q P(x, L) d x$.  Assuming the
FSS hypothesis, Eq.~\equ{eq:fss}, the $q$-th moment has the size
dependence:
\begin{equation}
  \left<x^q\right>_L = L^{\beta_x(q+1-\tau_x)} \int y^{q-\tau_x} {\cal
  F}(y) d y \sim 
  L^{\beta_x(q+1-\tau_x)},
  \label{eq:moment}
\end{equation}
where we have introduced the transformation $y=x/L^{\beta_x}$. In
general, one has $\left<x^q\right>_L \sim L^{\sigma_x(q)}$, where the
exponent $\sigma_x(q)$ can be obtained as the slope of the log-log
plot of $\left<x^q\right>_L$ as a function of $L$. If the FSS
hypothesis is indeed correct, we expect $\sigma_x(q) \sim
\beta_x(q+1-\tau_x)$, and therefore one can compute $\beta_x = d
\sigma_x(q) / d q$. For very small values of $q$ this is not correct,
since the integral in \equ{eq:moment} is dominated by its lower
cut-off.  Once computed the exponent $\beta_x$, the corresponding
$\tau_x$ is obtained using the scaling relation
$\sigma_x(1)=\beta_x(2-\tau_x)$.

In order to apply the moment analysis technique, we have performed
simulations of the P-Z and S-Z models in $d=2$, for sizes ranging from
$L=128$ to $L=1024$.  Statistical distributions are obtained by
averaging over $5 \times 10^6$ nonzero avalanches. As a consistency
check of our algorithm, we have estimated the average energy of the
system in the stationary state, defined by $\bar{E} = \left< \sum_i
  E_i \right> /L^2 $, where the brackets denote an average over at
least $10^6$ stable configurations. Extrapolating to an infinite
system size, we obtain a value $\bar{E}=0.630\pm0.005$ for the P-Z
model and $\bar{E}=0.626\pm0.005$ for the S-Z model, both in good
agreement with previous estimates $(\bar{E}\sim0.62-0.63)$
\cite{zhang89,giacometti98}.

We have computed the moments $\sigma_x(q)$ for our data from both the
P-Z and S-Z models. In the presence of FSS, as we have argued above,
one should expect to observe $d \sigma_x(q) / d q = \beta_x \equiv$
const.  Simulations on the Manna model, Ref.~\cite{lubeck00}, show
that this is indeed the case, with a slope for the moments
$\sigma_x(q)$ that reaches a saturation value for relatively small
values of $q$~\cite{noteq}.

\begin{figure}[t]
  \centerline{\epsfig{file=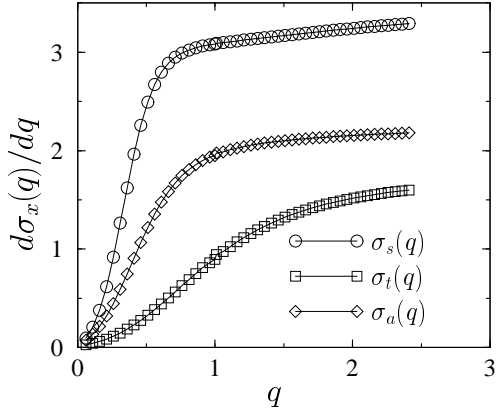, width=6.5cm}}
  \caption{Derivative of the exponents $\sigma_x(q)$ for the parallel Zhang
    model. The monotonous increase of the exponent indicates the lack
    of scaling in this model.}
  \label{fig:sigmazhang}
\end{figure}

\begin{figure}[t]
  \centerline{\epsfig{file=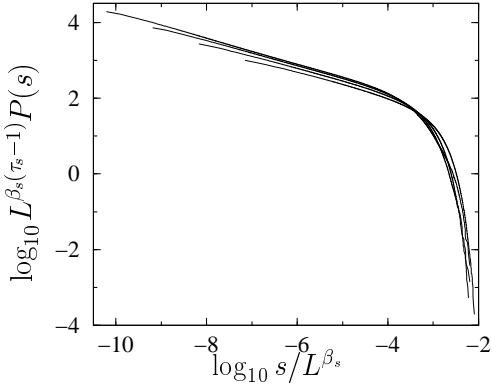, width=6.5cm}}
  \caption{Data collapse analysis of the integrated distribution of
    sizes, for the parallel Zhang model. System sizes
  are $L=128, 256, 512$, and $1024$. Exponents are  $\beta_s=3.39$ and
  $\tau_s=1.42$.} 
  \label{fig:fsszhang}
\end{figure}

In Fig.~\ref{fig:sigmazhang} we have plotted $d \sigma_x(q) / d q$ for
the P-Z model as a function of $q$, for $q<2.5$.  We observe that all
the moments present a noticeable curvature for all $q$, and do not
seem to reach a constant slope for the values of $q$ considered. For
values of $q$ larger than $4$, the slope of the moment functions seems
to finally achieve a saturation value.  As an estimate of this trend,
in Table~\ref{tablegaps} we report the value of the local slope at
different $q$'s, \cite{tebaldi99}, defined by $\Delta\sigma_x(q) =
\sigma_x(q+1)-\sigma_x(q)$.  Assuming that the curvature of the
moments at small $q$ is merely a crossover effect and that the real
exponents $\beta_x$ are given by the saturation plateaus at large $q$,
we are led to the values $\beta_s=3.39$, $\beta_t=1.74$, and
$\beta_a=2.25$. In particular, the value of $\beta_a$ is completely
unphysical: The maximum area of an avalanche cannot grow faster than
$L^d$ in $d$ dimensions, and thus $\beta_a$ must be smaller than $2$.
This tells us that the FSS form used in Eq.~(\ref{eq:moment}) is not
adequate in the case of the P-Z model, and leads to spurious results.
In what respects to the size and time distributions, we can check the
validity of this result by means of a data collapse technique: If the
FSS ansatz Eq.~\equ{eq:fss} is correct, then by rescaling $x \to
x/L^{\beta_x}$ and $P(x) \to L^{\beta_x \tau_x} P(x)$, we should
obtain distributions that collapse onto the same universal curve for
different values of $L$.  In Fig.~\ref{fig:fsszhang} we plot the data
collapse of the size distribution $P(s)$, with the exponents
$\beta_s=3.39$ and $\tau_s=1.42$,.  The really poor collapse of the
curves testifies that also the avalanche size distribution does not
fulfill the FSS in the P-Z model. We observe a similar lack of
collapse for the area distribution.

\begin{table}[b]
\begin{tabular}{cccccc}
$q$    &  $1$   &  $2$   &  $3$    &  $4$   & $5$ \\
\hline
$\Delta\sigma_s(q)$& $3.17$ & $3.30$ & $3.37$  & $3.39$ & $3.40$\\  
$\Delta\sigma_t(q)$& $1.29$ & $1.61$ & $1.70$  & $1.73$ & $1.74$\\  
$\Delta\sigma_a(q)$& $2.09$ & $2.19$ & $2.23$  & $2.25$ & $2.26$\\  
\end{tabular}
\caption{Local slope of the moment exponents $\sigma_x(q)$ in the
  parallel Zhang model, for different values of $q$.}
\label{tablegaps}
\end{table}

From Figs.~\ref{fig:sigmazhang} and \ref{fig:fsszhang}, and
Table~\ref{tablegaps}, we conclude that the Zhang model with parallel
updating violates the FSS hypothesis, Eq.~\equ{eq:fss}. Noticeably,
this lack of FSS is observed also in numerical simulations of the
directed version of the Zhang model \cite{alexei}.  We have also tried
to fit our data to the more general multifractal scaling form proposed
by Kadanoff {\em et al.} \cite{kad89}, and applied to the BTW model in
Ref.~\cite{tebaldi99}. In this form of multifractal analysis, one
tries to collapse the data to the form $\log (P(x,L)) / \log(\alpha
L)$ as a function of $\log (x) / \log(\alpha L)$, for a suitably
chosen constant $\alpha$. We have checked that this sort of scaling is
also not compatible with our data of the original Zhang model. In
particular, we have not succeeded in finding a constant $\alpha$ for
which the scaling is correct.

\begin{figure}[t]
  \centerline{\epsfig{file=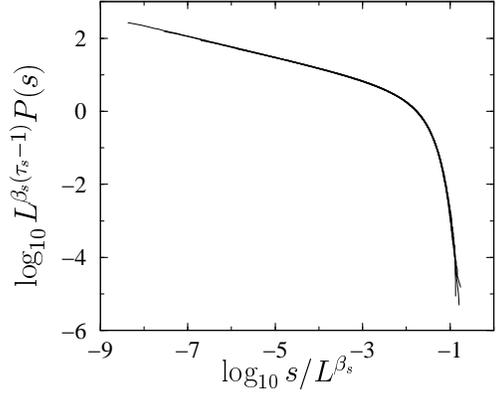, width=6.5cm}}
  \caption{Data collapse analysis of the integrated distribution of
    sizes for the sequential  Zhang model.
    System sizes are $L=128, 256, 512$, and $1024$. Exponents are
    $\tau_s=1.29$ and $\beta_s=2.78$.}
  \label{fig:fsszhangsequential}
\end{figure}

The moment analysis of the S-Z model yields good
results for the size and area distributions, with derivatives of the
moments $\sigma_x(q)$ that reach a saturation plateau for small values
of $q$. The values obtained are $\tau_s=1.29(2)$ and $\beta_s=2.78(2)$ 
for the size exponents, 
and in Fig.~\ref{fig:fsszhangsequential} we plot the
data collapse analysis for the size distribution. The perfect collapse
of this figure confirms the good scaling of this model. The same is 
obtained with the area distributions with exponents 
$\tau_a=1.43(2)$ and $\beta_s=1.94(2)$.
The time
distribution, however, shows a lack of scaling,
with a not clear trend for the derivative $\beta_t = d \sigma_t(q) / d q$ as
a function of $q$. This feature could have several origins included the 
possibility that in this case we do not have yet reached the scaling
regime because of the different time updating, that gives a very small
range of time scales.

The complex quality of scaling in the parallel Zhang model can be
attributed to the deterministic nature of its dynamical rules, which
is somehow smoothed by the stochastic updating in the sequential
model. In order to check this conjecture, we propose a variant of the
P-Z model, the stochastic parallel Zhang model (SP-Z), in which energy
is stochastically redistributed. The model is defined by the following
modifications of the relaxation rules. In the SP-Z model, an active
site $k$ loses all its energy, $E_k \to 0$, which is randomly
redistributed among its nearest neighbors. In the practical
implementation of this rule, we draw four random numbers
$\varepsilon_{k'}$, $0 \leq \varepsilon_{k'} \leq 1$, with $\sum_{k'}
\varepsilon_{k'} = 1$ and update the nearest neighbors $k'$ by $E_{k'}
\to E_{k'}+ \varepsilon_{k'} E_k$ \cite{note2}.  In this model, the
stochastic energy input is a random variable uniformly distributed in
$[0, \delta_{\rm max}]$, with $\delta_{\rm max}=0.25$.  Sites still
have a continuous spectrum of energy, but the new dynamical rules are
stochastic. If the assumption that stochasticity yields a standard
scaling behavior, then this model should be regarded as the continuous
counterpart of the original Manna model \cite{manna}.

\begin{figure}[t]
  \centerline{\epsfig{file=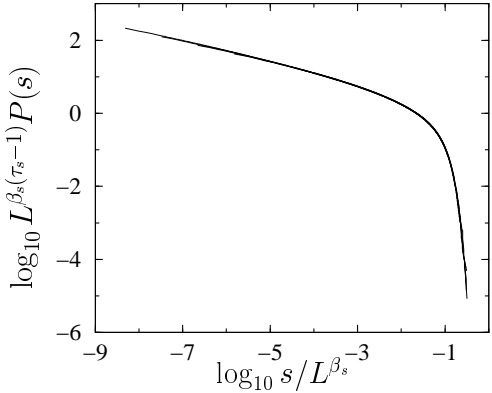, width=6.5cm}}
  \caption{Data collapse analysis of the integrated distribution of
    sizes for the stochastic parallel  Zhang model.
    System sizes are $L=128, 256, 512$, and $1024$. Exponents from
    Table~\protect\ref{tableexponents}.}
  \label{fig:fsszhangstochastic}
\end{figure}

\begin{table}[b]
\begin{tabular}{lcccccc}
Model   & $\tau_s$ & $\beta_s$ & $\tau_t$ & $\beta_t$ & $\tau_a$  &
$\beta_a$ \\ 
\hline
SP-Z & $1.28(1)$ & $2.76(1)$ & $1.50(2)$ & $1.53(2)$ & $1.35(1)$ & 
$2.02(2)$ \\
M   & $1.28(1)$ & $2.76(1)$ & $1.48(2)$ & $1.55(1)$ & $1.35(1)$ & 
$2.02(2)$ 
\end{tabular}
\caption{Critical exponents for 
  the  stochastic parallel Zhang model (SP-Z) and the 
  Manna (M) model. Figures in parenthesis indicate the statistical 
  uncertainty in the last digit. Data for the Manna model from 
  Refs.~\protect\cite{granada,lubeck00}. }
\label{tableexponents}
\end{table}

We have performed numerical simulations of the SP-Z model in system
sizes ranging from $L=128$ to $L=1024$, averaging the probability
distributions over $5 \times 10^6$ nonzero avalanches.  We observe
that slope of $\sigma_x(q)$ reaches a saturation value for very small
values of $q$, for sizes, areas, and also times.  In
Table~\ref{tableexponents} we report the values obtained for the
exponents $\beta_x$ and $\tau_x$. As expected, the exponents are in
perfect agreement with the values found in the Manna model. This fact
confirms the presence of a unique universality class for all stochastic 
models. As a final check, we show in Fig.~\ref{fig:fsszhangstochastic} the
data collapse analysis for the distribution of sizes.  The perfect
collapse of these plots should be compared with the poor result
obtained in the original Zhang model, shown in
Fig.~\ref{fig:fsszhang}.

In summary, applying the moment analysis technique, we have shown that
the scaling of the Zhang model depends on the updating rules
implemented in the simulations. The Zhang model with parallel update
violates the FSS hypothesis Eq.~\equ{eq:fss} for the avalanche
distributions of sizes, times, and areas.  The anomalous scaling is
stronger than in the BTW model, since data cannot be fitted even to
the more general multifractal scaling form. This, contrary to previous
claims~\cite{milshtein98,lubeck97b,giacometti98}, makes impossible to
assign any precise universality class to this model.  The sequential
updating introduces a small amount of stochasticity in the Zhang model
that, in this case, shows FSS for the size and area distributions. 
In spite of this property, the lack of scaling for the time distribution
does not allow to place the sequential Zhang model in a definite 
universality class.  The
anomalous scaling of the original Zhang model can be therefore
ascribed to the deterministic nature of the dynamical rules defining
the model. A stochastic versions of the model show a standard FSS
behavior, compatible with the universality class of the Manna model.

This work has been supported by the European Network under Contract
No.~ERBFMRXCT980183. We thank A. Stella, A. V\'{a}zquez, and
S. Zapperi for helpful comments and discussions.

\newpage


\begin{thebibliography}{10}


\bibitem{jensen98}
H.~J. Jensen, {\em Self-Organized Criticality} (Cambridge University
Press,  Cambridge, 1998).

\bibitem{btw1}
P. Bak, C. Tang, and K. Wiesenfeld, Phys. Rev. Lett. {\bf 59},  381
(1987). 

\bibitem{manna}
S.~S. Manna, J. Phys. A {\bf 24},  L363  (1991).

\bibitem{dhar}
D. Dhar, Physica A {\bf 263},  4  (1999).

\bibitem{milshtein98}
E. Milshtein, O. Biham, and S. Solomon, Phys. Rev. E {\bf 58},  303
(1998). 

\bibitem{granada}
A. Chessa, A. Vespignani, and S. Zapperi, Comput. Phys. Commun. {\bf
  121-122},  299  (1999).

\bibitem{lubeck00}
S. L\"{u}beck, Phys. Rev. E {\bf 61},  204  (2000).

\bibitem{vdmz99}
A. Vespignani, R. Dickman, M.~A. Mu{\~n}oz, and S. Zapperi,
Phys. Rev. E {\bf 62},  4564  (2000).

\bibitem{zhang89}
Y.-C. Zhang, Phys. Rev. Lett. {\bf 63},  470  (1989).

\bibitem{lubeck97b}
S. L{\"u}beck, Phys. Rev. E {\bf 56},  1590  (1997).


\bibitem{giacometti98}
A. Giacometti and A. {D{\'\i}az-Guilera}, Phys. Rev. E {\bf 58},  247
(1998). 

\bibitem{men}
M. {De Menech}, A.~L. Stella, and C. Tebaldi, Phys. Rev. E {\bf 58},
R2677  (1998).

\bibitem{tebaldi99}
C. Tebaldi, M. {De Menech}, and A.~L. Stella, Phys. Rev. Lett {\bf
  83},  3952  (1999).

\bibitem{kad89}
L.~P. Kadanoff, S.~R. Nagel, L. Wu, and S. Zhou, Phys. Rev. A {\bf
  39},  6524  (1989).

\bibitem{note1} 
In the original Zhang model \cite{zhang89} the energy
  increment $\delta$ in constant; the random increment used in this
  work leads to the same critical behavior, but renders the algorithm
  faster \cite{giacometti98}.


\bibitem{vz}
        G. Grinstein, in 
        {\it Scale Invariance, Interfaces and Nonequilibrium Dynamics}, 
        {\it NATO Advanced Study Institute, Series B: Physics},
        vol. 344, A. McKane et al., Eds. 
        (Plenum, New York, 1995);
        A. Vespignani and S. Zapperi, 
        Phys. Rev. E {\bf 57}, 6345  (1998).

\bibitem{cardy88}
 {\em Finite Size Scaling}, Vol.~2 of {\em Current Physics-Sources and 
  Comments}, edited by J.~L. Cardy (North Holland, Amsterdam, 1988).

\bibitem{notezhang}
In our nomenclature, the exponent $\beta_s$ is the fractal dimension
$D$ as  defined in other publications, while $\beta_t$ represents the
dynamic  critical exponent $z$.


\bibitem{noteq} When computing the moments, one has to be aware of the
limitations of the validity of Eq.~\equ{eq:moment} for large $q$. At
large $q$, the maximum contribution in the average \equ{eq:moment}
corresponds to the largest avalanches, of which one has the poorest
statistics. Therefore, the moment's estimation for large $q$ is
affected by large statistical errors.

\bibitem{alexei}
        A. V\'{a}zquez, cond-mat/0003420.


\bibitem{note2} 
Note the essential difference between this model and the Manna model
in the limit of infinite threshold recently introduce by L\"{u}beck
[S. L{\"u}beck, cond-mat\-/0008304].

\end{thebibliography}
\end{document}